\documentclass[twocolumn]{webofc}

\usepackage[varg]{txfonts}   
\usepackage{hyperref}
\usepackage{url}
\hypersetup{colorlinks=true,citecolor=blue,urlcolor=blue,linkcolor=blue}
\begin{document}
\title{Activities of the Korea ALICE group for the development and production of the next-generation silicon tracker}
%
%

\author{Sanghoon Lim\inst{1}\fnsep\thanks{\email{shlim@pusan.ac.kr}} \textit{for the ALICE Collaboration}}

\institute{Department of Physics, Pusan National University, Busan, South Korea 
}

\abstract{ALICE 3 is the proposed next-generation heavy-ion experiment at the CERN Large Hadron Collider (LHC), envisioned for operation during Run 5. The tracking system of ALICE 3 will consist of a high-precision vertex detector integrated into a retractable structure inside the beam pipe, complemented by a large-area outer tracker covering a broad pseudorapidity range. Both systems will be based on the Monolithic Active Pixel Sensor (MAPS) technology, building upon the developments realized for the recently upgraded ALICE Inner Tracking System (ITS2) and the future ITS3 upgrade. The total silicon area of the ALICE 3 tracking system is expected to be approximately five times larger than that of ITS2, presenting significant challenges in terms of large-scale sensor testing and module production. To address these challenges, research and development activities have been initiated in Korea, including the adaptation of an automated die-attach machine—commonly used in the semiconductor packaging industry—for efficient sensor-to-substrate assembly. This contribution presents the ongoing efforts of the Korea ALICE group toward the development and production of the ALICE 3 silicon tracker. The scope includes sensor evaluation, automated assembly techniques, and prototype module construction, with the aim of establishing scalable procedures for future mass production.
}
\maketitle
\section{Introduction}
\label{intro}
The ALICE (A Large Ion Collider Experiment) collaboration at CERN is advancing toward its next major upgrade—ALICE 3, planned for LHC Run 5, with the primary goal of conducting high-precision studies of the quark-gluon plasma (QGP) at unprecedented sensitivity levels~\cite{RefA, RefB}. A central component of this new detector is a state-of-the-art silicon tracking system based on Monolithic Active Pixel Sensors (MAPS), engineered to meet the stringent requirements of high spatial resolution, minimal material budget, fast time response, and radiation tolerance in the high-luminosity environment of future heavy-ion collisions.

\begin{figure}[h]
\centering
\includegraphics[width=0.48\textwidth,clip]{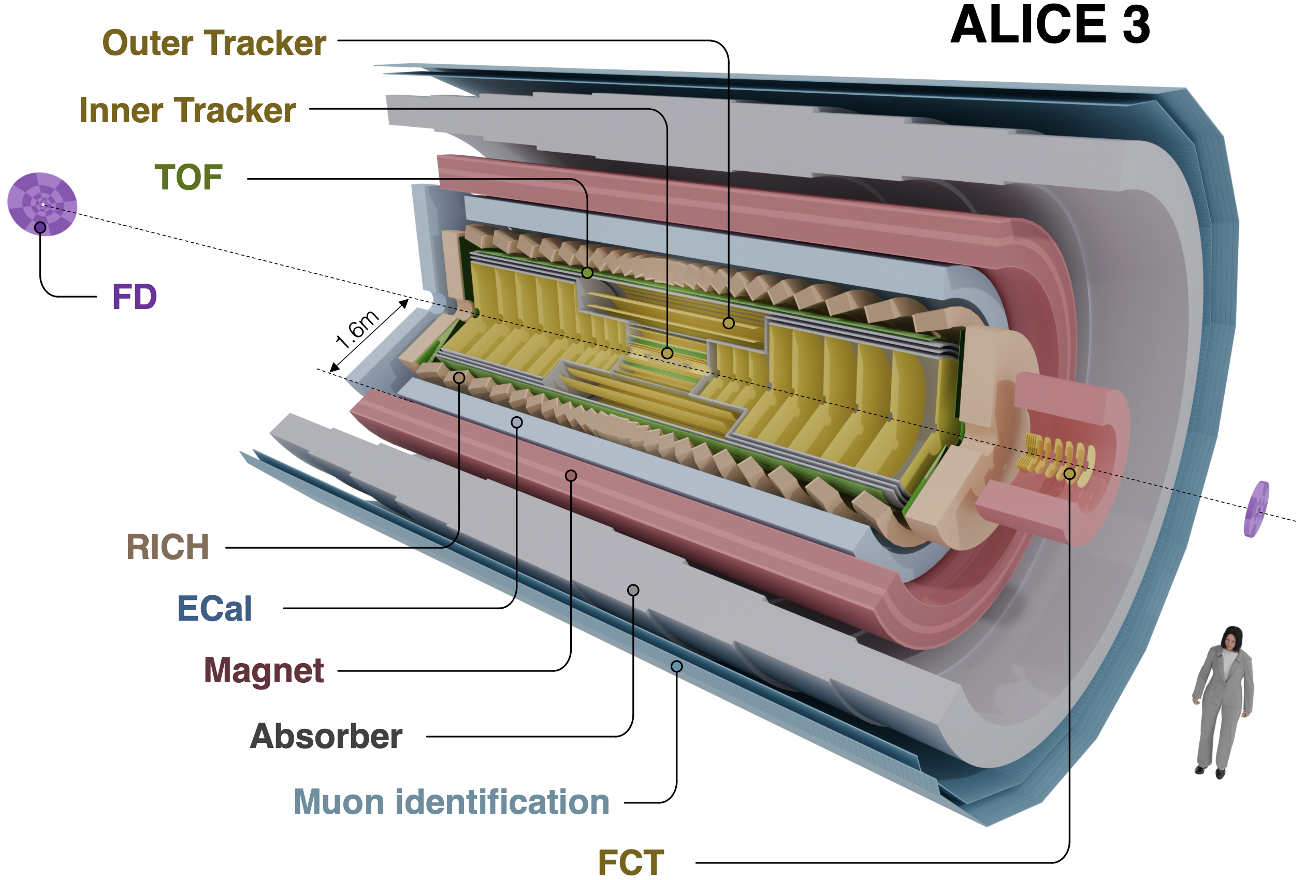}
\caption{The schematic view of the ALICE 3 detector.}
\label{fig-ALICE3}       
\end{figure}

As illustrated in Fig.~\ref{fig-ALICE3}, the ALICE 3 apparatus will feature a fully integrated suite of detectors optimized for low-$p_\mathrm{T}$ tracking and comprehensive particle identification. The detector system includes the Inner and Outer Trackers (both MAPS-based), inner and outer Time-of-Flight (TOF) detectors, Ring Imaging Cherenkov (RICH) detectors, a superconducting magnet, an electromagnetic calorimeter (ECal), a muon identifier (MID) with an integrated hadron absorber, and the Forward Conversion Tracker (FCT) placed within a small dipole magnet. Together, these components will enable ALICE 3 to conduct detailed investigations of QGP properties with enhanced precision and a broader physics reach compared to its predecessors.

The ALICE 3 tracking system is designed to meet the stringent performance requirements of reconstructing low-transverse-momentum ($p_{\mathrm{T}}$) particles in the extreme environment of high-multiplicity heavy-ion collisions. The Inner Tracker (IT) delivers unprecedented impact parameter resolution and vertexing performance, which are indispensable for the identification of multi-charm baryons and low-$p_{\mathrm{T}}$ beauty hadrons. Complementing this, the Outer Tracker (OT) extends the overall lever arm of the tracking system, significantly improving momentum resolution and enhancing the robustness of pattern recognition algorithms—both essential for ensuring reconstruction quality across the full detector acceptance.

The OT barrel geometry comprises four cylindrical layers positioned at radial distances of 30, 45, 60, and 80 cm. This configuration has been chosen based on detailed simulation studies that demonstrate the necessity of all four layers to preserve tracking performance in both low and high-$p_{\mathrm{T}}$ regimes. Attempts to reduce the number of barrel layers to three—for example, adopting a layout with layers at 30, 50, and 80 cm—were shown to result in a substantial degradation in transverse momentum resolution, especially in the absence of the outermost hit. In contrast, the four-layer design limits the resolution loss to within 20--25\%, even in cases of localized failure, thus validating its selection as the baseline layout for the ALICE 3 OT. To further enhance its physics reach, ALICE 3 incorporates forward tracking disks that extend the pseudorapidity coverage up to $|\eta| = 4$, surpassing the acceptance of previous ALICE configurations. This design ensures nearly continuous geometric coverage and enables long-range correlation measurements with large $\eta$-gaps and sufficient statistical precision. 

The OT utilizes Monolithic Active Pixel Sensors (MAPS) developed in the 65-nm CMOS imaging process by Tower Partners Semiconductor Co. (TPSCo), continuing the technological advancements of the ITS3 project. Each sensor has dimensions of approximately $3.2 \times 2.5~\mathrm{cm}^2$, with an additional peripheral region of $3.2 \times 0.15~\mathrm{cm}^2$ along its long edge reserved for readout electronics. A single 30-cm wafer accommodates 57 such sensors. Considering a production yield of approximately 67\%, the full-scale fabrication will require about 1300 wafers to produce 73,900 sensors for the barrel and 620 wafers to deliver 35,300 sensors for the endcap disks. For the estimates of the numbers of modules and for the R\&D on module assembly, a sensor size of $3.2 \times 2.5~\mathrm{cm}^2$ is currently considered.

The sensor modules for the OT are standardized to support both barrel and endcap implementations. Each module comprises eight sensors arranged in two rows with 200 µm spacing, covering an active area of $12.88 \times 5.04~\mathrm{cm}^2$, or approximately 65 cm². These sensors are bonded to a flexible printed circuit (FPC) and subsequently attached to a high-thermal-conductivity substrate designed with mechanical locators for accurate placement. In total, approximately 5,620 modules are required for the OT barrel, and 2,688 modules are needed to equip 12 replicated endcap disks. Accounting for production yield (85\%) and additional units for pre-production and validation, the total number of modules to be fabricated reaches approximately 10,750. Given the large production scale and the need for timely delivery, the module assembly process will be outsourced to industry, and preliminary engagements with commercial vendors have already been initiated.

\section{Activities of the Korea ALICE group}

The Korean ALICE group has participated in the ITS2 project by contributing to ALPIDE sensor testing and the production of Outer Barrel (OB) modules \cite{RefC, RefD}, thereby gaining essential experience in silicon detector assembly. The project was successfully executed through collaborations with domestic semiconductor-related companies, including FUREX, C-ON Tech, and MEMSPACK. Building on this experience, the group aims to contribute to the ALICE 3 Outer Tracker (OT) development by promoting the automation and industrialization of module production.

During the ITS2 OB production, module assembly was carried out in parallel at five different sites using the ALICIA machine~\cite{RefE}. The precise placement of sensors onto assembly jigs was automated using chip-handling equipment. However, subsequent steps—such as glue dispensing on the Flexible Printed Circuit (FPC) and the attachment of the sensor to the FPC—were performed manually. The adhesive used, Araldite 2011, is a two-component epoxy that requires pre-mixing and a long curing time of approximately five hours, limiting daily production to about two modules per site. In the case of ALICE 3 OT, the number of required modules is approximately five times greater than that of the ITS2 OB. Therefore, significant improvement throughout the production is critical to meet the schedule. The Korean team plans to leverage its prior experience and existing industrial partnerships to enhance production efficiency through greater automation and process optimization.

\subsection{Industrialization of module assembly}

The Korean ALICE group collaborated with MEMSPACK, which participated in the wire bonding process during the ITS2 OB module production, to explore automation techniques for module assembly. General-purpose die-bonder machines used in industry are equipped with advanced vision systems and chip-handling mechanisms, enabling precise operations such as glue dispensing on PCBs, sensor alignment, and attachment. At MEMSPACK, equipment such as the MRSI-705 and Datacone evo 2200+ systems is available and has been utilized for this purpose~\cite{RefF}. Using these machines, dummy modules based on the ITS2 OB module design were successfully produced, demonstrating the feasibility of automating key stages in the detector module assembly process.

\begin{figure}[h]
\centering
\includegraphics[width=0.48\textwidth,clip]{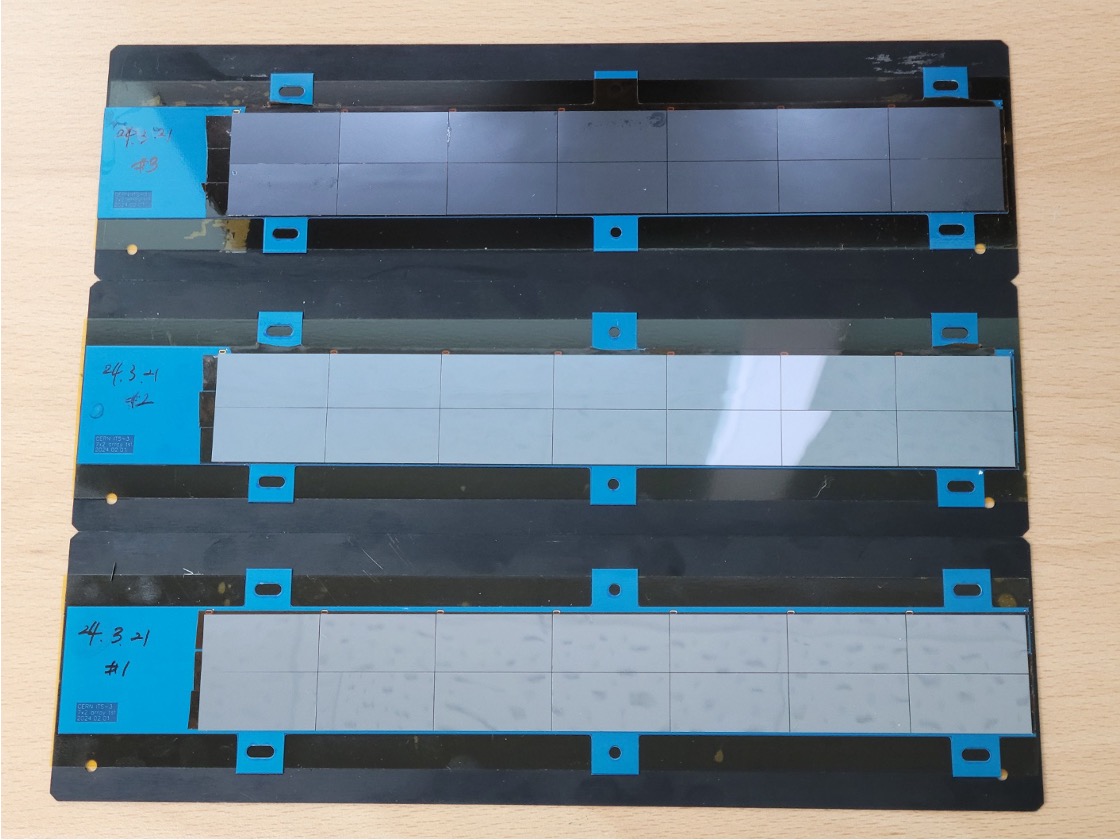}
\caption{Dummy modules assembled with the ITS2 Outer Barrel (OB) mechanical design.}
\label{fig-dummy1}       
\end{figure}

Figure~\ref{fig-dummy1} shows a dummy module based on the ITS2 OB design, produced at MEMSPACK using 100 $\mu$m-thick silicon sensors and 300 $\mu$m-thick rigid PCBs. While these components are thicker than those used in actual detector modules, they were intentionally selected to validate the placement accuracy of a general-purpose die-bonder machine and to optimize the procedure for placing sensors onto PCBs. In this setup, the PCB is first positioned on an assembly jig. The die-bonder’s chip-handling tool then picks up a sensor from the tray and places it onto the PCB by recognizing alignment markers on both the sensor and the PCB, thus ensuring high-precision placement. In the initial tests, to reduce variability and focus on placement accuracy, the sensor and PCB were joined using double-sided tape instead of glue. The positional precision was evaluated by measuring the inter-sensor spacing, and it was found that the gap between adjacent sensors was consistently maintained within 150 $\mu$m in both X and Y directions, demonstrating excellent uniformity in placement accuracy.

\begin{figure}[h]
\centering
\includegraphics[width=0.48\textwidth,clip]{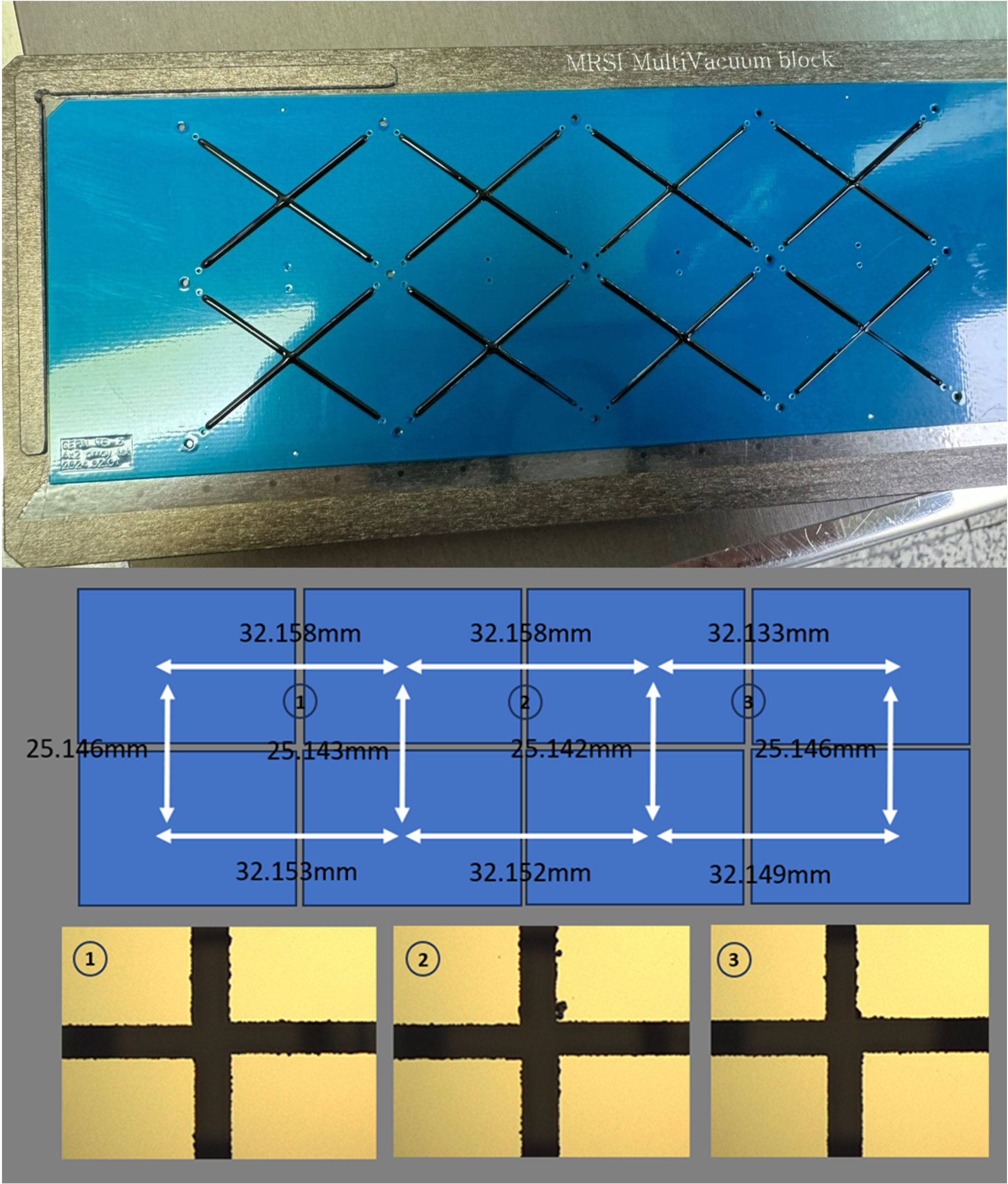}
\caption{Dummy modules assembled with the ALICE 3 OT module mechanical design.}
\label{fig-dummy2}       
\end{figure}

In the next step, module assembly was performed using epoxy adhesives, with dummy chips and PCBs corresponding to the ALICE 3 OT module design ($3.2 \times 2.5~\mathrm{cm}^{2}$). Two types of epoxy were tested. The first was a heat-cured epoxy: as shown in the top panel of Fig.~\ref{fig-dummy2}, epoxy was dispensed onto the PCB, the sensor was positioned, and the assembled module was then transferred to a curing station, where it was cured at approximately 100 $^\circ$C for 30 minutes. The second type was a two-component epoxy that can be cured either at room temperature or with heating. For this test, room-temperature curing was used. The bottom panel of Fig.~\ref{fig-dummy2} shows the measurement of the gap between sensors immediately after placement. As in the previous tests, the inter-sensor spacing was found to be highly uniform, confirming the consistency and precision of the placement process.

\begin{figure}[h]
\centering
\includegraphics[width=0.48\textwidth,clip]{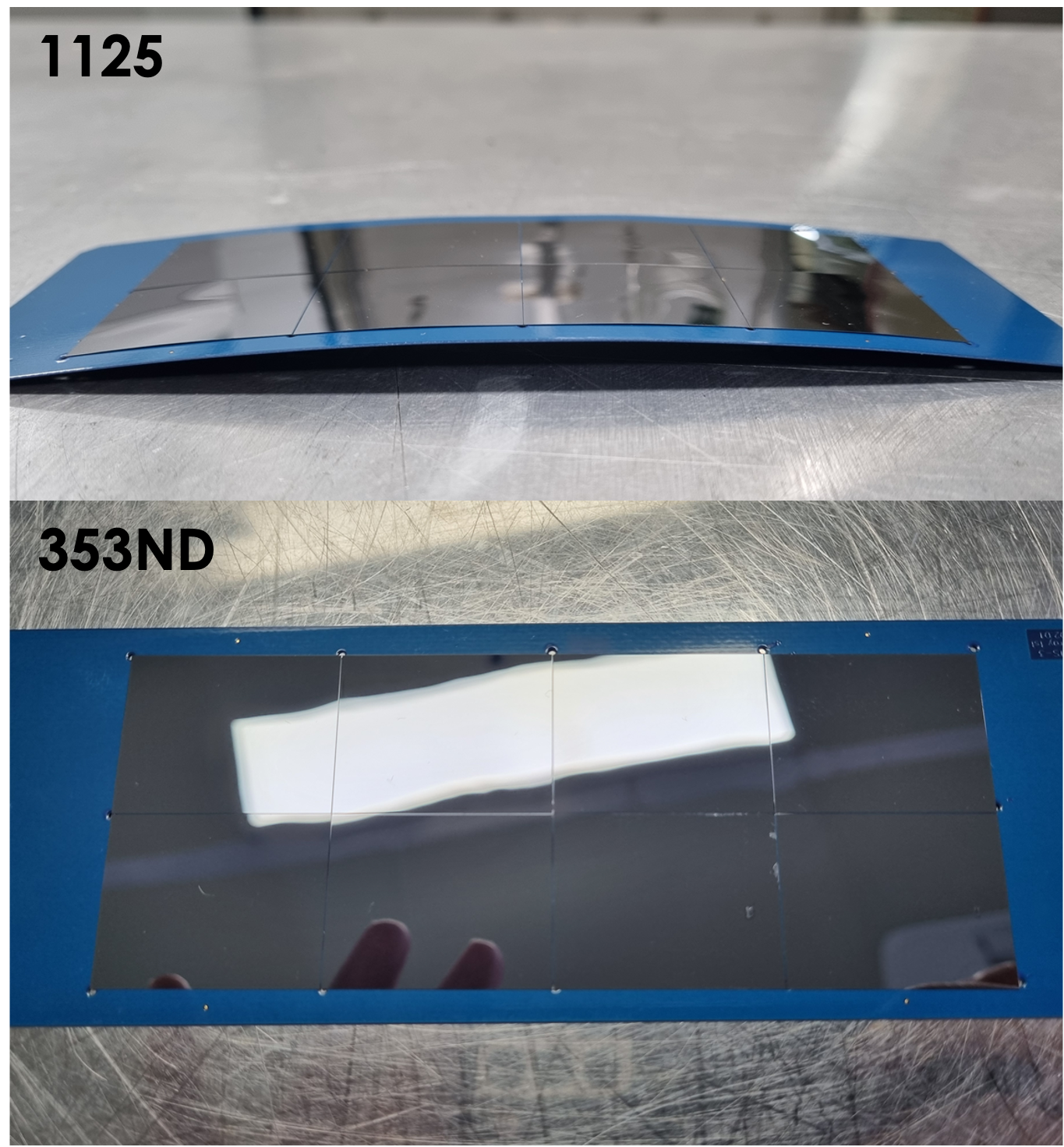}
\caption{Dummy modules assembled with two different glues.}
\label{fig-dummy3}       
\end{figure}

Figure~\ref{fig-dummy3} shows the dummy modules after the curing process. The top panel corresponds to the heat-cured epoxy, while the bottom panel shows the module cured at room temperature using a two-component epoxy. In the case of heat curing, a deformation in the module shape was observed when viewed from the side, indicating that the sensor positions had become warped during the process. In contrast, the module cured at room temperature exhibited no visible physical deformation. Measurements taken before and after curing confirmed that the sensor positions remained uniform, demonstrating that room-temperature curing preserved the mechanical stability of the module.

\begin{figure}[h]
\centering
\includegraphics[width=0.48\textwidth,clip]{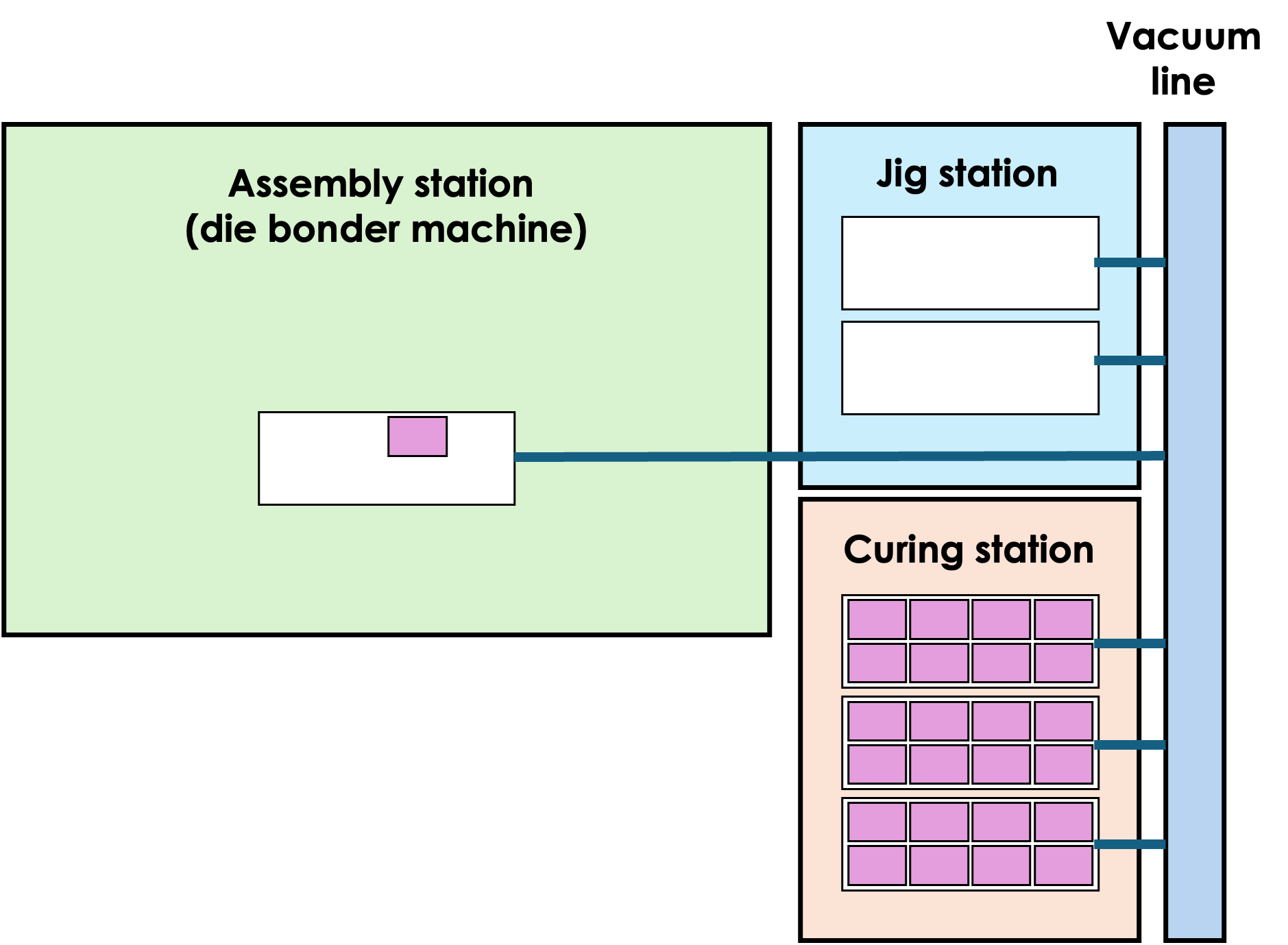}
\caption{Conceptual diagram of the automated setup for continuous production of detector modules.}
\label{fig-prodline}       
\end{figure}

Through this process, we successfully validated the positioning accuracy using a general-purpose die-bonder. The next goal is to establish and verify a setup for continuous module production. Figure~\ref{fig-prodline} shows a schematic layout of this setup. On the right side, multiple assembly jigs are connected to a vacuum line. After sensors are attached to the PCBs using the assembly machine, the jig, while maintaining vacuum, is transferred to a curing station. As the curing process proceeds, the next jig is used to assemble the following module, enabling a continuous production workflow. This method is expected to improve production efficiency, with an estimated completion time of approximately 10 minutes per single module.

\subsection{Glue radiation hardness test}

\begin{figure}[h]
\centering
\includegraphics[width=0.48\textwidth,clip]{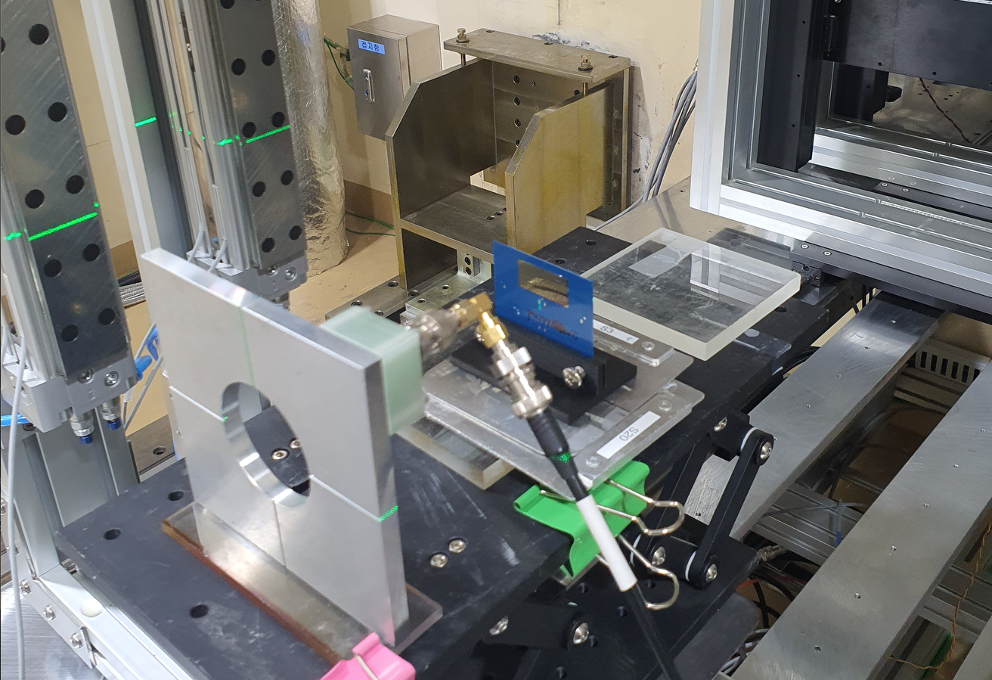}
\caption{Setup for evaluating glue radiation hardness under proton exposure at KOMAC.}
\label{fig-komac}       
\end{figure}

The type of glue used during the module production process has a significant impact on both productivity and quality, primarily due to factors such as ease of handling and curing time. In the ALICE ITS2 project, Araldite 2011 was selected for its radiation hardness, although it requires a long curing time of more than five hours. To improve productivity in future module production, a study is currently underway to identify an optimal adhesive by evaluating the radiation tolerance of several candidate glues that are easier to handle during assembly.

At MEMSPACK, several dummy samples were prepared by attaching a single dummy sensor to a PCB using different types of glue. These samples were then irradiated with a 20 MeV proton beam at KOMAC, followed by mechanical testing to evaluate the post-irradiation integrity of the glue bonds. In the first test campaign, three types of glue were evaluated under a radiation exposure of approximately 2.5 Mrad total ionizing dose (TID) and a fluence of $1.5\times 10^{13}$ 1 MeV neq/cm$^{2}$--ten times the expected radiation dose for the ALICE 3 Outer Tracker.

\begin{figure}[h]
\centering
\includegraphics[width=0.48\textwidth,clip]{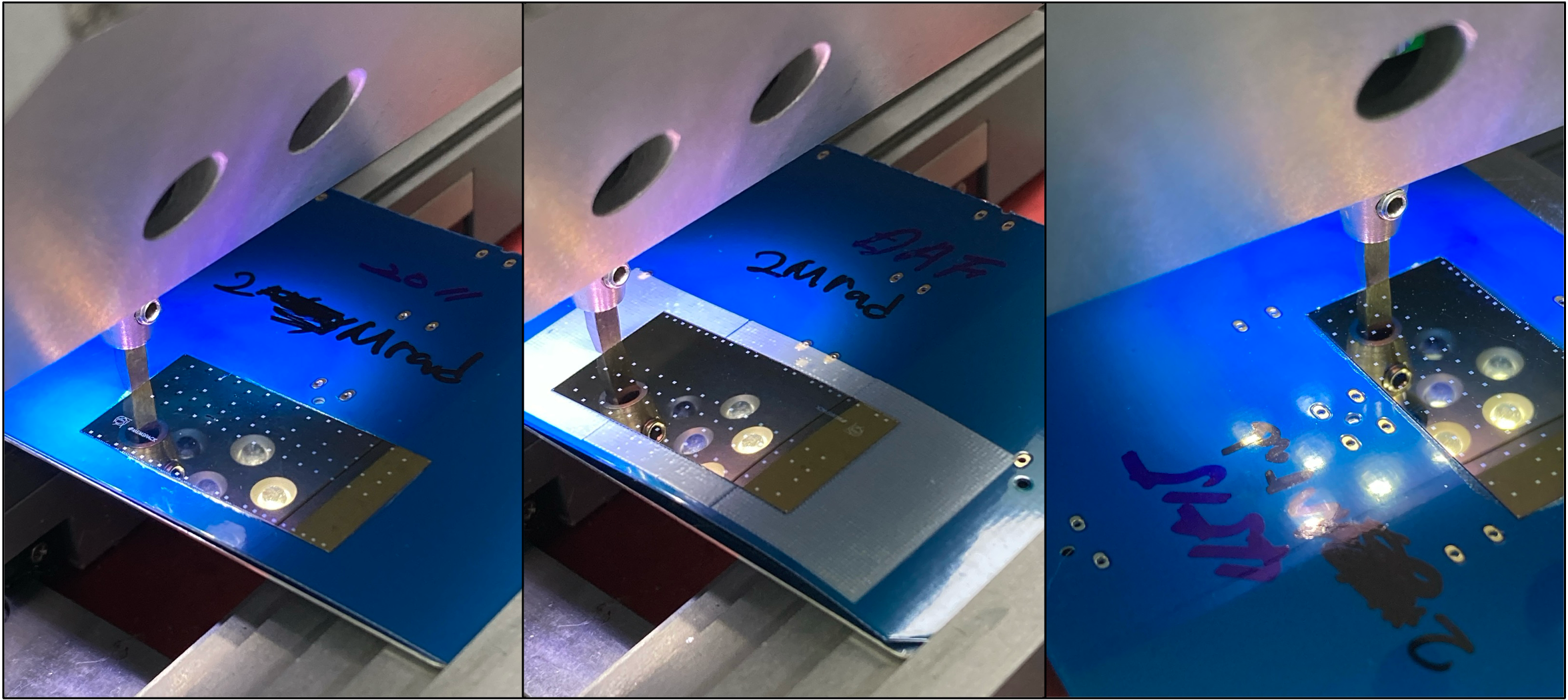}
\caption{Procedure for performing mechanical testing using ASTM-F1269 equipment.}
\label{fig-test}       
\end{figure}

Figure~\ref{fig-komac} shows the setup used for irradiation at KOMAC. After irradiation, a shear test was conducted to assess whether the sensors remained securely attached to the PCB. As shown in Fig.~\ref{fig-test}, a mechanical tip applied a force of 8.8 kgf to push the sensor off the board. In this first round of testing, all glues passed the shear test. Further testing is planned to include a more refined validation procedure, utilizing a broader range of samples that explore various glue types and dispensing volumes. These efforts aim to identify the most suitable adhesive for large-scale module production in the ALICE 3 Outer Tracker.

\section{Summary}

In preparation for the ALICE 3 experiment at the CERN LHC, the Korea ALICE group is actively contributing to the development of the next-generation silicon tracking system, focusing on scalable module production techniques. Building on its experience from the ALICE ITS2 project, the team has initiated R\&D efforts to automate and industrialize the assembly of MAPS-based detector modules, collaborating with industrial partners such as MEMSPACK. Using general-purpose die-bonder machines, the group successfully demonstrated high-precision sensor placement and verified uniform inter-sensor alignment with both adhesive and tape-based bonding methods. In parallel, efforts to identify optimal adhesives for mass production are underway at KOMAC, utilizing 20 MeV proton beams for radiation hardness and mechanical shear testing. These activities aim to validate room-temperature-curable glues for improved productivity while maintaining radiation tolerance. 

%
%
%

\end{document}